\documentstyle[12pt]{article}
\renewcommand{\thefootnote}{\fnsymbol{footnote}}
\catcode`\@=11
\def\makepreprititle{\par
  \begingroup
  \def\thefootnote{\fnsymbol{footnote}}
  \def\
@makefnmark{\hbox 
  to 0pt{$^{\@thefnmark}$\hss}} 
  \if@twocolumn 
  \twocolumn[\@makepreprititle] 
  \else \newpage
  \global\@topnum\z@ 
  \@makepreprititle \fi\thispagestyle{empty}\@thanks
  \endgroup
  \setcounter{footnote}{0}
  \let\makepreprititle\relax
  \let\@makepreprititle\relax
  \gdef\@thanks{}\gdef\@author{}\gdef\@title{}
  \gdef\@preprintnumber{}\gdef\@preprintdate{}\gdef\subtitle{}
  \let\thanks\relax}
\def\preprintnumber#1{\gdef\@preprintnumber{#1}}
\def\preprintdate#1{\gdef\@preprintdate{#1}}
\def\subtitle#1{\gdef\@subtitle{#1}}
\def\@makepreprititle{\newpage
{\def\baselinestretch{1}
  \begin{flushright} \@preprintnumber \par
  \@preprintdate \end{flushright} } \par
  \begin{center}
\vskip 1.5em
  {\LARGE \@title \par} \vskip 2.5em 
  {\large \lineskip .5em
  \begin{tabular}[t]{c}\@author 
  \end{tabular}\par}
  \vskip 1em {\large \@date} \end{center}
  \par
  \vfil} 
\date{\sl Department of Physics, Tohoku University\\Sendai, 980 Japan}
\preprintdate{~}
\preprintnumber{~}
\subtitle{}
\def\abstract{\if@twocolumn
\section*{Abstract}
\else \normalsize 
\begin{center}
{\bf Abstract\vspace{-.5em}\vspace{0pt}} 
\end{center}
\quotation 
\addtocounter{page}{-1}
\fi}
\def\endabstract{\if@twocolumn\else\endquotation\fi}
\catcode`\@=12

\preprintnumber{DPSU-9303\\TU-439}
\preprintdate{June, 1993}
\title{Probing Symmetry-Breaking Pattern\\Using Sfermion Masses}
\author{\Large
Yoshiharu~Kawamura\\
\large\sl
Department of Physics, Shinshu University\\
\large\sl Matsumoto, 390 Japan\\
\\
\rm\Large Hitoshi Murayama\thanks{Present address:
Theoretical Physics Group, Lawrence Berkeley Laboratory,
University of California, CA 94720.}~
 and  Masahiro Yamaguchi\\
\large \sl Department of Physics, Tohoku University\\
\large \sl Sendai, 980 Japan}
\date{~}
\begin{document}
\makepreprititle
\begin{abstract}
We study the mass spectrum of superparticles within supersymmetric
grand unified models. For gaugino masses, it is pointed out that the
GUT-relation in the $SU(5)$ model is applicable to a more general case
where a grand-unified gauge group breaks down to the standard model
gauge group by several steps.  We also show that the mass spectrum of
squarks and sleptons carries the information on the breaking pattern
of the gauge symmetry. It is demonstrated in some $SO(10)$ models how
the scalar mass spectrum distinguishes various $SO(10)$ breaking
patterns from each other.
\end{abstract}
\newpage

The Grand Unified Theory (GUT) has been attractive as a promising
framework to explain the law of nature since its proposal 
\cite{Georgi-Glashow}. The huge difference between the GUT scale and
the weak scale gives rise to the famous gauge hierarchy problem. The
prime  motivation of introducing
supersymmetry (SUSY) in a GUT \cite{SUSYGUT} is that it will give 
 a partial solution
to this problem:  SUSY can stabilize  
the hierarchy between the GUT scale and the weak scale
against radiative corrections \cite{naturalness}.
    
This
wonderful theoretical framework, the SUSY-GUT, is consistent with   the
precise measurement of the weak-scale gauge coupling  
constants at
LEP \cite{LEP} for the minimal particle content of SUSY standard model
\cite{Amaldi}.  Furthermore, the present
non-observation of the nucleon decay is shown to be still consistent
with the minimal version of the SUSY-GUT \cite{dimen5}.
It is, however, not clear from the LEP data alone whether the minimal
version of the SUSY-GUT is the whole story. For example, the solar neutrino
experiments \cite{solar} suggest the neutrino oscillation \`a la
Mikheyev--Smirnov--Wolfenstein \cite{MSW}, and it is naturally
incorporated into the $SO(10)$ grand unification with seesaw mechanism
\cite{seesaw}. While the direct breaking of the $SO(10)$ group into the
standard model group $G_{SM} = SU(3)_C \times SU(2)_L \times U(1)_Y$ is
obviously consistent with the LEP data, there are possibilities that
there is an intermediate scale with chain symmetry breaking.\footnote{An
explicit example was first given in 
Ref.~\cite{Deshpande} from different motivation.}

An important virtue of the SUSY-models is that the soft SUSY-breaking
mass parameters can be novel probes of physics at very high energies.
In this letter we point out that the gaugino mass spectrum generally
satisfies the  GUT-relation  as far as the
standard model gauge group is embedded into a simple group,
irrespective of the symmetry breaking pattern. 
On the other hand,
the squark and slepton mass spectrum will be shown to carry the
information on the breaking pattern of the gauge symmetry. Therefore,
the gaugino and the scalar mass spectrum will play a complementary role
to select among the models of SUSY-GUT experimentally. We will
demonstrate how the scalar mass spectrum distinguishes various $SO(10)$
breaking patterns from each other.

We  first 
consider the gaugino mass spectrum.  We expect that the gaugino mass
parameters are  common at the unification scale $M_U$. 
Though it is known that the vacuum expectation values of the fields
responsible for the GUT symmetry breaking would give a non-universal
contribution to  the gaugino
masses \cite{non-universal}, it 
is suppressed by powers of $M_U /M_{ Planck}$, which can be
neglected as far as the unification scale is not very close to the
Planck scale.  Then we can show that the spectrum of the gaugino
masses  always satisfies the so-called GUT-relation
\cite{Inoue}\footnote{We define
$\alpha_2 = \alpha/\sin^2 \theta_W$ and $\alpha_1 = \frac{5}{3}
\alpha/\cos^2 \theta_W$ throughout the paper. The symbols $M_i$ stand
for gaugino masses of various gauge groups, $g_i$ the gauge coupling
constants, and $\alpha_i = g_i^2/4\pi$. The GUT-relation is a
consequence of the one-loop renormalization group equations. The
 GUT-relation fails to hold at the two-loop level \cite{two-loop}, 
 but numerically the
effect is small \cite{Yamada}.  The threshold effects are also
neglected since the threshold corrections on the gaugino masses at
the GUT-scale can be almost absorbed in the threshold corrections on the
gauge coupling constants, so that there are no large logarithms
appearing in the corrections to the 
GUT-relation of the gaugino masses \cite{GHM}. The
GUT-relation can be violated only at the order of $\alpha/\pi$.}
\begin{equation}
\frac{M_1 (m_Z)}{\alpha_1 (m_Z)}
	= \frac{M_2 (m_Z)}{\alpha_2 (m_Z)}
	= \frac{M_3 (m_Z)}{\alpha_3 (m_Z)}
	= \frac{M_{GUT} (M_U)}{\alpha_{GUT} (M_U)},
		\label{GUT-relation}
\end{equation}
irrespective of the breaking pattern if the gauge group is unified in a
simple group at a high mass scale $M_U$. 
Though one can prove (\ref{GUT-relation}) in
a general framework, we will demonstrate here that the
GUT-relation indeed holds in some breaking patterns of $SO(10)$ group,
namely the direct breaking into the standard model gauge group (``direct
breaking''),
\newcommand{\verylongrightarrow}{
\relbar\joinrel\relbar\joinrel\relbar\joinrel\rightarrow}
\newcommand{\breaksto}[1]{\mathop{\verylongrightarrow}\limits^{#1}}
\begin{eqnarray}
SO(10) \breaksto{M_U} G_{SM},
\end{eqnarray}
``Pati-Salam'',
\begin{eqnarray}
SO(10) \breaksto{M_U} SU(4)_{PS} \times SU(2)_L \times SU(2)_R 
	 \breaksto{M_{PS}} G_{SM},
\end{eqnarray}
and ``3221'' \cite{Deshpande}
\begin{eqnarray}
SO(10) \breaksto{M_U}
	SU(3)_C \times SU(2)_L \times SU(2)_R \times U(1)_{B-L} 
	\breaksto{M_{B-L}} G_{SM}.
\end{eqnarray}
We assume the particle content of the Minimal Supersymmetric Standard
Model (MSSM) below the breaking scales, $M_U$, $M_{PS}$ or $M_{B-L}$,
for each breaking patterns respectively. These breaking scales are
supposed to be much higher than the SUSY-breaking scale.

It is obvious that the GUT-relation of the gaugino masses holds in the
``direct'' breaking. For the ``Pati-Salam'' case, the $SU(3)_C$ gaugino
comes solely from $SU(4)_{PS}$ gaugino, and $SU(2)_L$ remains unbroken
at the symmetry-breaking scale of the ``Pati-Salam'' symmetry ($M_{PS}$). 
Then the following equalities hold:
\begin{equation}
\frac{M_3 (m_Z)}{\alpha_3 (m_Z)} 
	= \frac{M_4 (M_{PS})}{\alpha_4 (M_{PS})}
	= \frac{M_{10} (M_U)}{\alpha_{10} (M_U)}	\label{M3PS}
\end{equation}
for the $SU(3)_C$ gaugino, and
\begin{equation}
\frac{M_2 (m_Z)}{\alpha_2 (m_Z)} 
	= \frac{M_{2L} (M_{PS})}{\alpha_{2L} (M_{PS})}
	= \frac{M_{10} (M_U)}{\alpha_{10} (M_U)}	\label{M2PS}
\end{equation}
for the $SU(2)_L$ gaugino. Here, $M_{10}$ represents the $SO(10)$ gaugino
mass.  There is a complication for the $U(1)_Y$ gaugino because it is a
mixture of $SU(2)_R$ and $SU(4)_{PS}$ gauginos. Since
$\sqrt{\frac{3}{5}} Y =
\sqrt{\frac{2}{5}} T^{15}_{4} +
\sqrt{\frac{3}{5}} T^3_{2R}$,\footnote{We adopt the notation $T^{15}_{4}
= \frac{1}{\sqrt{24}} {\rm diag} (1, 1, 1, -3)$ and $T^3_{2R} =
\frac{1}{2} {\rm diag} (1, -1)$.} the gauge coupling constants satisfy
\begin{equation}
	\frac{1}{\alpha_1} = \frac{2}{5} \frac{1}{\alpha_4}
		+ \frac{3}{5} \frac{1}{\alpha_{2R}}
\end{equation}
at $M_{PS}$. On the other hand, the gauge fields $A^\mu$ mix as
$g_1^{-1} A_1 
= g_4^{-1} \sqrt{\frac{2}{5}} A_{4}^{15} + g_{2R}^{-1}
\sqrt{\frac{3}{5}} A_{2R}^3$, and the gaugino fields $\lambda$ mix
correspondingly as
\begin{equation}
	\frac{1}{g_1} \lambda_1 
	= \frac{1}{g_4} \sqrt{\frac{2}{5}} \lambda_{4}^{15}
	+ \frac{1}{g_{2R}} \sqrt{\frac{3}{5}} \lambda_{2R}^3,
\end{equation}
as required from the supersymmetry. 
Thus the gaugino mass is given as
\begin{equation}
\frac{M_1(m_Z)}{\alpha_1(m_Z)} 
	= \frac{M_1(M_{PS})}{\alpha_1(M_{PS})} 
	= \frac{2}{5} \frac{M_4(M_{PS})}{\alpha_4(M_{PS})} 
	+ \frac{3}{5} \frac{M_{2R}(M_{PS})}{\alpha_{2R}(M_{PS})} 
	= \frac{M_{10}(M_U)}{\alpha_{10}(M_U)} ,
					\label{M1PS}
\end{equation}
where we used the solution to the renormalization group equations of
$M_4$ and $M_{2R}$. Therefore from
Eqs.~(\ref{M3PS},\ref{M2PS},\ref{M1PS}), the gaugino masses $M_3$, $M_2$
and $M_1$ satisfy the GUT-relation~(\ref{GUT-relation}). Exactly
the same argument applies to the ``3221'' breaking pattern as well.

Summarizing the above discussion, we have demonstrated that the
gaugino masses satisfy the GUT-relation even with an intermediate
scale, irrespective of the breaking patterns, as far as the whole
gauge groups are unified in $SO(10)$.  More general treatment will be
given elsewhere.\footnote{Note that our proof of the GUT-relation does
not depend on the particle content of the models.}

An immediate consequence of the above observation is the following.
The measurement of the gaugino masses at the weak scale will give us a
useful suggestion on whether the standard model gauge group $G_{SM}$ is
embedded into a simple group or not, irrespective of the breaking
pattern. Recall that the gaugino masses do not satisfy the
GUT-relation~(\ref{GUT-relation}) in general if the gauge group is not
unified as in flipped $SU(5)$ model.  However, one cannot distinguish
between the models which have different breaking patterns but from the
same unification group.

Now we discuss the renormalization group evolution of the squark and
slepton masses in the three breaking patterns. 
The renormalization-group equations for the scalar masses are given by
\begin{eqnarray}
      \frac{d}{d \ln \mu}m_a^2 (\mu)
   & = & -\frac{2}{\pi} \sum_i C_2 (R_i^a) \alpha_i (\mu) M_i (\mu)^2
       + \frac{3}{10 \pi} Y_a \alpha_1 (\mu) S (\mu), 
\label{rg1} \\
      \frac{d}{d \ln \mu} S (\mu)
   & = & \frac{b_1}{2 \pi} \alpha_1 (\mu) S(\mu),
\label{rg2} \\
     S& =& \sum_a Y_a n_a m_a^2 \label{S}
\end{eqnarray}
where $i$ represents the gauge group, $a$ the species of the scalar, 
$C_2 (R_i^a)$ the second Casimir invariant of the gauge group $i$ for
the species $a$, $Y_a$ the hypercharge, and $n_a$ the multiplicity of
the species $a$. 
In Eq.~(\ref{rg1}) we have neglected the Yukawa coupling contribution.
This approximation should be valid for the first- and the
second-generation fields. It is straightforward to generalize our
results to the third generation by considering the effects of Yukawa
coupling contributions.   The contribution from $S$ is usually ignored since 
it is absent  under the assumption of the universal scalar mass.
For the MSSM, it is\footnote{
We refer
to the chiral multiplets as $q$ for left-handed quark, $l$ left-handed
lepton, $u$ right-handed up, $d$ right-handed down and $e$ for
right-handed charged lepton. The tilde represents their scalar
component.  $m^2_1$ and $m^2_2$ stand for the soft SUSY breaking mass
terms of the Higgs bosons with hypercharge $-1/2$ and $+1/2$, respectively.}
\begin{equation}
    S=m_2^2 - m_1^2 +\sum_{generations}(m_{\tilde q}^2
              -2 m_{\tilde u}^2 +m_{\tilde e}^2-m_{\tilde l}^2
                +m_{\tilde d}^2).
\end{equation}
The coefficients of the beta function
$b_i$ are defined by
\begin{equation}
	\frac{d}{d \ln \mu} \alpha_i^{-1}
		= - \frac{b_i}{2\pi} .
\end{equation}
Solving the renormalization-group equations we  obtain
\begin{eqnarray}
    m_a^2 (\mu) & =& m_a^2 (\mu_0)
		- \sum_i \frac{2}{b_i} C_2 (R_i^a) 
			(M_i^2 (\mu) - M_i^2 (\mu_0))
\cr
        & &         +\frac{3}{5 b_1}Y_a (S(\mu)-S(\mu_0)) , \\
    S(\mu) &=& \frac{\alpha_1 (\mu)}{\alpha_1 (\mu_0)} S(\mu_0).
\end{eqnarray}
Since the quantity $S$ at the weak scale can be determined through
measurements, we can easily take its contribution to the scalar masses
into account.  Therefore the appearance of $S$ in the above equations
does not prevent us from going further.

Let us examine how the sfermion mass spectrum at the breaking scale
$M_{SB}$ reflects the pattern of the gauge symmetry breaking.  One may
naively expect that scalars belonging to a single multiplet above
$M_{SB}$ have a common mass at $M_{SB}$. 
 There is, however, an
important complication due to the presence of the so-called $D$-term
contribution to the scalar masses which appears when the rank of the
gauge group is reduced. In Refs.~\cite{Hagelin}, it was demonstrated
that the $D$-term contribution occurs when the gauge symmetry is
broken at an intermediate scale due to the soft SUSY breaking terms.
The existence of the $D$-term contribution in a more general situation
was suggested in Ref.~\cite{Faraggi}.  One can show a sizable $D$-term
contribution generally exists once the soft SUSY breaking terms in the
scalar potential are not universal \cite{KMY}.\footnote{
The assumption that the scalar masses have universal structure
is a strong one. In fact, it is known that the non-universal
soft SUSY breaking parameters emerge in the effective theory
derived from superstring \cite{LK}. Even if they are
universal at the Planck scale as in minimal supergravity or
SUSY-breaking in dilaton $F$-term, the radiative corrections
between the Planck and the symmetry breaking scale generally induce
non-universality of the scalar masses.}  Then we obtain the
correction to the scalar mass terms of the form
\begin{equation}
   \sum_{\alpha} g_{\alpha}^2 \langle D^{\alpha} \rangle 
       \varphi^{\dagger} T^{\alpha}  \varphi.
\end{equation}  

A $D$-term can be non-zero if the corresponding broken generator 
 commutes with all unbroken generators.  Such broken generators
constitute a subgroup $G'$. In particular,  the $D$-term contribution
arises when the rank of the group is reduced due to the gauge symmetry
breaking.  When $SO(10)$ breaks to $G_{SM}$, the rank is 
reduced by one and $G'$ is just $U(1)$. Thus the
$D$-term contribution is expressed by one parameter $D$.
This is also the case for chain breaking of $SO(10)$.

In the ``direct'' breaking, the scalar masses satisfy\footnote{ The  
normalization and the sign of $D$ are arbitrary.}
\begin{eqnarray}
m_{\tilde{q}}^2 &=& m_{16}^2 + g_{10}^2 D,\\
m_{\tilde{u}}^2 &=& m_{16}^2 + g_{10}^2 D,\\
m_{\tilde{e}}^2 &=& m_{16}^2 + g_{10}^2 D,\\
m_{\tilde{l}}^2 &=& m_{16}^2 - 3 g_{10}^2 D,\\
m_{\tilde{d}}^2 &=& m_{16}^2 - 3 g_{10}^2 D,
\end{eqnarray}
at the $SO(10)$ unification scale $M_U$. Since $M_U$ can be determined
from the renormalization group equations of gauge coupling constants
as well as the gaugino masses, the only free parameters are $m_{16}^2$
and $D$. Having five measurable scalar masses,\footnote{It is probably
impossible to measure the SUSY-breaking part of the right-handed
sneutrino mass as far as it has SUSY-invariant mass of the
intermediate scale.} we can solve for $m_{16}^2$ and $D$, and still
have three relations among them.  A typical evolution of the scalar
masses below $M_U$ is depicted in Fig.~1. It is clear that we can
check whether the measured scalar masses are consistent with the
``direct'' breaking.  Note that $S$ at the scale $M_U$ is
\begin{equation}
   S(M_U)=m_2^2-m_1^2.
\end{equation}
If the two Higgs doublets belong to the same {\bf 10} representation,
the above equation becomes 
\begin{equation}
   S(M_U)=-4g_{10}^2 D.
\end{equation}
In a more complicated model, we do not have predictability on $S$.
However, the quantity $S$ can be measured at the
weak scale, so that one can easily incorporate $S$ into analysis without
knowing the physics at the GUT-scale. 
In Fig.~1, we took $S=0$ for definiteness.

For the ``Pati-Salam'' case, we have four parameters: two independent
scalar masses $m_L^2$ for the left-handed multiplet and $m_R^2$ for the
right-handed multiplet below $M_U$, the parameter $D$, and the scale
$M_{PS}$ itself. Thus we can solve for them from $m_{\tilde{q}}$,
$m_{\tilde{l}}$, $m_{\tilde{u}}$, $m_{\tilde{d}}$ and $m_{\tilde{e}}$,
and still have one relation among them. The scalar masses satisfy
\begin{eqnarray}
m_{\tilde{q}}^2 &=& m_L^2 + g_4^2 D,\\
m_{\tilde{u}}^2 &=& m_R^2 - (g_4^2 - 2 g_{2R}^2) D,\\
m_{\tilde{e}}^2 &=& m_R^2 + (3 g_4^2 - 2 g_{2R}^2) D,\\
m_{\tilde{l}}^2 &=& m_L^2 - 3 g_4^2 D,\\
m_{\tilde{d}}^2 &=& m_R^2 - (g_4^2 + 2 g_{2R}^2) D,
\end{eqnarray}
at $M_{PS}$. Recall that $g_4$ and $g_{2R}$ can be computed from the
weak-scale coupling constants as a function of $M_{PS}$ alone. After
eliminating $D$, $m_L^2$ and $m_R^2$, we obtain two relations,
\begin{eqnarray}
m_{\tilde{q}}^2 (M_{PS}) - m_{\tilde{l}}^2 (M_{PS}) 
	&=& m_{\tilde{e}}^2 (M_{PS}) - m_{\tilde{d}}^2 (M_{PS}),
		\label{PS1} \\
g_{2R}^2 (M_{PS}) (m_{\tilde{q}}^2 - m_{\tilde{l}}^2) (M_{PS}) 
	&=& g_4^2 (M_{PS}) (m_{\tilde{u}}^2 - m_{\tilde{d}}^2) (M_{PS}),
		\label{PS2}
\end{eqnarray}
and one of them should be used to determine $M_{PS}$. A typical
evolution of the scalar masses is shown in Fig.~2.\footnote{Here we have
assumed the presence of two chiral superfields for each of the
representations ({\bf 4}, {\bf 1}, {\bf 2}), ({\bf 4}$^\ast$, {\bf 1},
{\bf 2}), ({\bf 1}, {\bf 1}, {\bf 3}) above $M_{PS}$ to achieve the
unification of the gauge coupling constants at the unification scale
$M_U$. The precise assignment of the quantum number is, however,
irrelevant to our discussion since we can test the relations (\ref{PS1})
and (\ref{PS2}) without specifying the beta-function above $M_{PS}$.}
The Fig.~2(b) provides a magnified view around $M_{PS}$ so that the
relations (\ref{PS1}) and (\ref{PS2}) are visible.
\setcounter{footnote}{0}

$S$ in the ``Pati-Salam'' case is
\begin{equation}
   S=m_2^2-m_1^2 +24 (g_4^2-g_{2R}^2) D
\end{equation}
at $M_{PS}$.  As in the case of the ``direct'' breaking, 
the value of $S$ depends on the
Higgs structure.  For the simplest case where the two Higgs doublets
are in a single ({\bf 1}, {\bf 2}, {\bf 2}), we have
\begin{equation}
   S=-4g_{2R}^2D +24 (g_4^2-g_{2R}^2) D.
\end{equation}
In Fig.~2, we took $S=0$ for definiteness as in the ``direct'' case.

The breaking pattern by ``3221'' has smaller symmetry below $M_U$ than
the ``Pati-Salam'' case, and hence there are more parameters. Above the
symmetry-breaking scale of $SU(3)_C \times SU(2)_L \times SU(2)_R \times
U(1)_{B-L}$ symmetry ($M_{B-L}$), there are four independent scalar
masses $m_{\tilde{q}L}^2$, $m_{\tilde{l}L}^2$, $m_{\tilde{q}R}^2$ and
$m_{\tilde{l}R}^2$ corresponding for left- and right-handed quark/lepton
doublets.  The scalar masses have the following contributions from the
$D$-terms,
\begin{eqnarray}
m_{\tilde{q}}^2 &=& m_{\tilde{q}L}^2 + g_{B-L}^2 D,\\
m_{\tilde{u}}^2 &=& m_{\tilde{q}R}^2 - (g_{B-L}^2 - 2 g_{2R}^2) D,\\
m_{\tilde{e}}^2 &=& m_{\tilde{l}R}^2 + (3 g_{B-L}^2 - 2 g_{2R}^2) D,\\
m_{\tilde{l}}^2 &=& m_{\tilde{l}L}^2 - 3 g_{B-L}^2 D,\\
m_{\tilde{d}}^2 &=& m_{\tilde{q}R}^2 - (g_{B-L}^2 + 2 g_{2R}^2) D.
\end{eqnarray}
We have six unknown parameters $m_{\tilde{q}L}$, $m_{\tilde{l}L}$,
$m_{\tilde{q}R}$, $m_{\tilde{l}R}$, $M_{B-L}$ and $D$, in contrast to
the five observables. Therefore, we cannot solve for these parameters
and neither check these relations. The scalar mass spectrum looks just
disordered. A typical evolution is depicted in Fig.~3.\footnote{We have
taken the same particle content as in Ref.~\cite{Deshpande}, with a
triplicate of representation ({\bf 1}, {\bf 1}, {\bf 2}, ${\bf
-3/\sqrt{24}}$) and its conjugate under $SU(3)_C \times SU(2)_L \times
SU(2)_R \times U(1)_{B-L}$ symmetry.}  Note that $S$ in this case is
rather complicated.   
In the figure, we have again took $S=0$ for definiteness.

In summary, we have shown that the gaugino masses and scalar masses
carry complementary information on the symmetry breaking of the
unification group. Assuming  the  universal gaugino masses
at the GUT-scale, the gaugino masses satisfy  the GUT-relation   even with
the chain breaking of the gauge symmetry, as far as the standard model
gauge group $SU(3)_C \times SU(2)_L \times U(1)_Y$ is embedded into a
simple group. The breaking pattern is irrelevant. Therefore, the gaugino
masses supply a unique tool to infer whether the standard model gauge
group is unified in a simple group or not.

On the other hand, the scalar masses carry the information on the
nearest symmetry breaking pattern above the weak-scale. For models
with relatively large gauge group like $SO(10)$ itself or Pati-Salam
group $SU(4)_{PS} \times SU(2)_L \times SU(2)_R$, we can solve for the
original parameters of the model, and can also check whether the
scalar mass spectrum is consistent with the model or not.  It is
remarkable that the relations are obtained without specifying the
particle contents above the  breaking scale.  For models
with relatively small gauge group, $SU(3)_C
\times SU(2)_L \times SU(2)_R \times U(1)_{B-L}$, one may not be able to
extract the original mass parameters. In any case, one can distinguish
among the models by measuring the scalar masses.

\newpage
\section*{Figure Captions}

\renewcommand{\labelenumi}{Fig.~\arabic{enumi}}
\begin{enumerate}
\item A typical evolution of the scalar masses under the ``direct'' breaking
of $SO(10) \rightarrow G_{SM}$. $S$ in Eq.~(\ref{S}) is taken to be
zero.
\item A typical evolution of the scalar masses under the chain
breaking $SO(10) \rightarrow SU(4)_{PS} \times SU(2)_L \times SU(2)_R
\rightarrow G_{SM}$. The $D$-term contributions to the scalar masses are
depicted in Fig.~2(b) so that the relations (\ref{PS1}) and (\ref{PS2}) are
visible. $S$ in Eq.~(\ref{S}) is taken to be zero.
\item A typical evolution of the scalar masses under the chain
breaking $SO(10) \rightarrow SU(3)_C \times SU(2)_L \times SU(2)_R
\times U(1)_{B-L} \rightarrow G_{SM}$. The $D$-term contributions to the
scalar masses are depicted in Fig.~3(b). $S$ in Eq.~(\ref{S}) 
is taken to be zero.
\end{enumerate}

\newpage
\begin{thebibliography}{99}
\bibitem{Georgi-Glashow} H.~Georgi and S.L.~Glashow, {\sl
Phys.~Rev.~Lett.} {\bf 32}, 438 (1974). 

\bibitem{SUSYGUT}
S.~Dimopoulos  and H.~Georgi, {\sl Nucl. Phys.}\/ {\bf B193}, 150
(1981);\\
N.~Sakai, {\sl Z. Phys.}\/ {\bf C11}, 153 (1981).
 
\bibitem{naturalness} 
M.J.G.~Veltman, {\sl Acta Phys. Pol.}\/ {\bf B12}, 437 (1981);\\
L.~Maiani, {\it Gif-sur-Yvette Summer School on Particle Physics},\/ 11th,
Gif-sur-Yvette, France, 1979 (Inst. Nat. Phys. Nucl. Phys. Particules, Paris,
1979);\\
S.~Dimopoulos and S.~Raby, {\sl Nucl. Phys.}\/ {\bf B192}, 353 (1981);\\
E.~Witten, {\sl Nucl. Phys.}\/ {\bf B188}, 513 (1981);\\
M.~Dine, W.~Fischler, and M.~Srednicki, {\sl Nucl. Phys.}\/ {\bf B189}, 575 
(1981).




\bibitem{LEP} LEP Collaborations, {\sl Phys. Lett.}\/ {\bf B276}, 247 (1992). 

\bibitem{Amaldi} 
P.~Langacker and M.-X.~Luo, {\sl Phys. Rev.}\/ {\bf D44}, 817
(1991);\\ 
U.~Amaldi, W.~de~Boer and H.~F\"{u}rstenau, {\sl Phys. Lett.}\/ {\bf
260B}, 447 (1991);\\ 
W.J.~Marciano, Brookhaven preprint, BNL-45999, April 1991.


\bibitem{dimen5} 
J.~Hisano, H.~Murayama, and T.~Yanagida, {\sl Phys. Rev. Lett.}\/ {\bf
69}, 1014 (1992); {\sl Nucl. Phys.}\/ {\bf B402}, 46 (1993)

\bibitem{solar} 
R.~Davis, invited talk presented at ``International Symposium on
Neutrino Astrophysics'', 19--22 October 1992, Takayama/Kamioka, Japan;\\
Kamiokande-II Collaboration, K.S.~Hirata {\it et al.}\/,
{\sl Phys. Rev.}\/ {\bf D44}, 2241 (1991); ERRATUM {\it ibid}\/. {\bf
D45}, 2170 (1992);\\
SAGE Collaboration, A.I.~Abazov {\it et al.}\/,
{\sl Phys. Rev. Lett.}\/ {\bf 67}, 3332 (1991);\\
GALLEX Collaboration, P.~Anselmann {\it et al.},\/ {\sl Phys. Lett.}\/
{\bf B281} (1992).

\bibitem{MSW} 
S.~Mikheyev and Y.~Smirnov, {\sl Nuovo Cimento}\/ {\bf 9C} (1986) 17;\\
L.~Wolfenstein, {\sl Phys. Rev.}\/ {\bf D17} (1978) 2369.

\bibitem{seesaw} 
T.~Yanagida, in {\sl Proceedings of the Workshop on Unified
Theory and Baryon Number of the Universe},\/ eds. O.~Sawada and A.~Sugamoto
(KEK, 1979) p.95;\\
M.~Gell-Mann, P.~Ramond and R.~Slansky, {\sl Supergravity},\/ eds. P.~van
Nieuwenhuizen and D.~Freedman (North Holland, Amsterdam, 1979).


\bibitem{non-universal}
J.~Ellis, K.~Enqvist, D.V.~Nanopoulos and K.~Tamvakis, {\sl Phys.
Lett.}\/ {\bf 155B}, 381 (1985);\\
M.~Drees, {\sl Phys. Lett.}\/ {\bf 158B}, 409 (1985); {\sl Phys.
Rev.}\/ {\bf D33}, 1468 (1986).
 
\bibitem{Inoue} 
K.~Inoue, A.~Kakuto, H.~Komatsu and S.~Takeshita, 
{\sl Prog. Theor. Phys.}\/ {\bf 68}, 927 (1982); 
{\sl ibid.}\/ {\bf 71}, 413 (1984). 

\bibitem{two-loop}
Y.~Yamada, {\sl Phys. Lett.}\/ {\bf B316}, 109 (1993); \\ 
S.P.~Martin and M.T.~Vaughn, Northeastern preprint,  NUB-3072-93TH,
August 1993. 

\bibitem{Yamada}
Y.~Yamada, KEK preprint, KEK-TH-371, August 1993.
 
\bibitem{GHM} T.~Goto, J.~Hisano, and H.~Murayama,
Tohoku University preprint, TU-423, June 1993, to appear in {\sl Phys.
Rev.}\/ {\bf D}.

\bibitem{Deshpande} N.G.~Deshpande, E.~Keith, and T.G.~Rizzo, 
{\sl Phys. Rev. Lett.}\/ {\bf 70}, 3189 (1993).

 
\bibitem{Hagelin} 
M.~Drees, {\sl Phys. Lett.}\/ {\bf 181B}, 279 (1986);\\
J.S.~Hagelin and S.~Kelley, {\sl Nucl. Phys.}\/ {\bf
B342}, 95 (1990).

\bibitem{Faraggi}
A.E.~Faraggi, J.S.~Hagelin, S.~Kelley, and D.V.~Nanopoulos, {\sl Phys.
Rev.}\/ {\bf D45} 3272 (1992).

\bibitem{KMY} Y.~Kawamura, H.~Murayama and M.~Yamaguchi, in
preparation. 

\bibitem{LK}
L.E.~Ib\'a\~nez and D.~L\"ust, {\sl Nucl. Phys.}\/ {\bf B382}, 305 (1992);
B.~de~Carlos, J.A.~Casas and C.~Mu\~noz, {\sl Phys. Lett.}\/ {\bf
B299}, 234 (1993); 
V.S.~Kaplunovsky  and J.~Louis, {\sl Phys. Lett.}\/ {\bf B306}, 269 (1993).


\end{thebibliography}
\end{document}